\documentclass[final,3p]{CSP}
\usepackage{amssymb}
\usepackage{changepage}
\usepackage{amsthm}

\newtheorem{proposition}{Proposition}

\usepackage{amssymb}
\usepackage{changepage}
\usepackage{algorithm}
\usepackage{algpseudocode}
\usepackage{amsthm}
\usepackage{marvosym}
\usepackage{color}

\usepackage[T1]{fontenc}
\usepackage[utf8]{inputenc}
\usepackage{mathtools}   

\usepackage{graphicx}
\usepackage{bm}
\usepackage{latexsym}
\usepackage{amsfonts}
\usepackage{amsmath}
\usepackage{amssymb}
\usepackage{yfonts}
\usepackage{amssymb}
\usepackage{pax}
\usepackage{wasysym}
\usepackage{graphicx}
\usepackage{bm}
\usepackage{latexsym}
\usepackage{amsfonts}
 
\usepackage{pax}
\usepackage{wasysym}

\begin{document}

\begin{frontmatter}

\title{Nonabelian multiplicative integration and curvature obstructions for surface holonomy}

\author[mymainaddress,secondaddress]{Hollis Williams}

\address[mymainaddress]{Department of Physics, University of Warwick, Coventry CV4 7AL, United Kingdom}

\address[secondaddress]{ Theoretical Sciences Visiting Program, Okinawa Institute of Science and 
Technology Graduate University, Onna 904-0495, Japan}

\fntext[fn1]{Current address: Department of Mathematics and Statistics, University of Exeter, Exeter EX4 4QF, UK}

\begin{abstract}\rm
\begin{adjustwidth}{2cm}{2cm}{\itshape\textbf{Abstract:}} 
Surface holonomy plays a central role in higher gauge theory, bundle gerbes and the geometric formulation of Wess--Zumino terms in string theory. In this work, we consider the relation between surface holonomy and nonabelian multiplicative integration on surfaces.  In this framework, we interpret the local Stokes law as a curvature obstruction law for higher holonomy and investigate its consequences in the abelian setting.  We derive a global three-dimensional Stokes relation and show that it reproduces the familiar Wess-Zumino phase formula. In particular, the phase difference between two surfaces with common boundary is governed by the integral of the corresponding $3$-form curvature over an interpolating three-manifold.  These results provide a geometric interpretation of multiplicative integration on surfaces in terms of surface holonomy and clarify its relationship with the classical theory of bundle gerbes and Wess-Zumino terms. We conclude by discussing possible extensions to nonabelian higher gauge theories and their relation to Wilson surface operators and generalized symmetries.

\end{adjustwidth}
\end{abstract}
\end{frontmatter}

\section{Introduction}

\noindent
It has been known for many years that parallel transport along a curve and the associated notion of holonomy can be generalized to higher-dimensional parallel transport along surfaces with an associated surface holonomy \cite{schreiber2,baez1}. 
The main physical motivation for studying higher-dimensional holonomy originates in two-dimensional conformal field theories (in particular, Wess-Zumino-Witten models), where the Wess-Zumino term in the worldsheet action has long been interpreted as a surface holonomy \cite{witten2,gaw1,schreiber1}.  
WZW models are among the most important two-dimensional rational conformal field theories, forming the building blocks for coset models in the coset construction~\cite{kapa,witten1}.  Several mathematical frameworks have been developed to formalize higher parallel transport and surface holonomy, including Deligne cohomology and bundle gerbes with connection and curving, the connective structures of Breen and Messing, and more recently, multiplicative gerbes and smooth 2-group extensions~\cite{deligne, murray, gad2, breen, gad1,bunk1}.

A homotopy-theoretic treatment of higher-dimensional parallel transport has been given by Baez, Schreiber and Waldorf in terms of 2-bundles with 2-connections, where parallel transport is described by smooth transport 2-functors ~\cite{baez1,schreiber3}.  Although categorical approaches are conceptually powerful, they rely on higher category theory and abstract constructions which are unfamiliar to most physicists and differential geometers.  An alternative, more concrete analytic framework was introduced by Yekutieli under the name of nonabelian multiplicative integration (abbreviated in the article as MI) \cite{yek}.  In this approach, nonabelian line and surface integrals are constructed as explicit limits of elementary Riemann products, similarly to ordinary path-ordered exponentials, but extended to higher dimensions \cite{chen}.  The theory provides well-defined, rigorous analytic control and convergence estimates for these limits, without using any machinery from category and homotopy theory.  Although originally developed in the context of twisted deformation quantisation, it turns out this formalism reproduces the basic structure of higher parallel transport as developed by Schreiber and Waldorf, but in a much more elementary setting \cite{yek2,yek3, bressler}.

In this article, we examine the local multiplicative Stokes theorem of Yekutieli from the perspective of higher parallel transport and surface holonomy, showing that the MI formalism naturally associates surface holonomies to crossed-module-valued $2$-connections and provides an analytic framework for studying curvature obstructions in higher gauge theory.  For a sufficiently small smooth $3$-simplex, this theorem expresses the product of the surface holonomies of its boundary faces in terms of the associated $3$-curvature.  Geometrically, this may be interpreted as a local curvature-obstruction law: the failure of surface holonomy to be invariant under infinitesimal deformations is governed by the corresponding curvature $3$-form.  Under an abelian assumption on the higher gauge field, the local multiplicative Stokes law globalizes to a three-dimensional Stokes relation. As a consequence, the difference between the surface holonomies associated to two surfaces with common boundary is determined by the exponential of the integrated $3$-curvature over an interpolating three-manifold. We show that this reproduces the familiar Wess-Zumino phase relation appearing in two-dimensional sigma models.

The relation between surface holonomy and the integral of a $3$-form curvature is already known in the abelian gerbe literature.  In particular, the holonomy of bundle gerbes with connection and curving  provides an elegant realization of the abelian Wess–Zumino phase \cite{murray, wess, carey, freed, waldorf}.  However, there still remains the case of how to generalise to the nonabelian case: nonabelian generalizations of gerbes do exist, but they are technically involved and usually formulated using higher category theory \cite{baez1,schreiber2,picken}.  Our perspective in this article is that it should be possible to derive a relatively explicit form for the nonabelian Wess-Zumino phase using the purely analytic theory of multiplicative integration.  Although we do not prove this directly, the derivation of the abelian phase in this framework provides strong evidence that this should be possible.  It is likely that it will be necessary to construct a global version of multiplicative integration on surfaces, which we will not attempt here.  This complete construction would enable interesting applications to nonabelian higher gauge theories and Wilson surface observables.  Several frameworks for nonabelian surface holonomy do exist, including transport 2-functors and nonabelian gerbes, but a fully analytic version is still an open problem.

The article is structured as follows.  In Section 2, we briefly outline background material from Yekutieli's theory of nonabelian multiplicative integration on surfaces, including notation and conventions.  In Section 3, we provide our main results on curvature obstruction and the Wess-Zumino phase.  This includes the derivation of the abelian phase law directly from multiplicative integration.  We also discuss the nonabelian generalization.  We finish with conclusions and outlook in Section 4.

 \section{Background}

 \subsection{Notation and conventions}

\noindent
Throughout this article, all manifolds, maps, and differential forms are
assumed to be of class $C^{\infty}$.  For each integer $n\ge0$, we denote by
$\Delta^n=\{(t_0,\dots,t_n)\in\mathbb{R}^{n+1}\mid
 t_i\ge0,\,\sum_i t_i=1\}$ the standard real simplex with vertices
$v_0,\dots,v_n$.
Given a smooth manifold $M$, a smooth $n$--simplex in $M$
is a smooth map $\sigma\colon\Delta^n\to M$.
Unless otherwise stated, all simplices are oriented in the standard
order of their vertices.  A path in~$M$ is a smooth map
$\sigma\colon\Delta^1\to M$ with initial and end points denoted by $\sigma(v_0)$ and $\sigma(v_1)$, respectively.  Given two paths $\sigma_1,\sigma_2$, their concatenation is
denoted $\sigma_2\!\ast\!\sigma_1$.
The inverse path $\bar\sigma$ is defined by reversal of the orientation
of~$\Delta^1$.
For surfaces, we use the notation
$\tau\colon\Delta^2\to M$.  The boundary $\partial\tau$ is the oriented loop obtained by traversing
the edges of~$\Delta^2$ in cyclic order.  When necessary, we will use the original terminology of Yekutieli and refer to the
pair $(\sigma,\tau)$ consisting of a path
$\sigma\colon\Delta^1\to M$ and a surface
$\tau\colon\Delta^2\to M$ with $\sigma(v_1)=\tau(v_0)$ as a kite \cite{yek}.
The point $\sigma(v_0)$ is called the base point of the kite.

Next, let $G$ and $H$ be Lie groups with Lie algebras
$\mathfrak g$ and $\mathfrak h$, respectively.
The exponential maps are written
$\exp_G\colon\mathfrak g\to G$ and
$\exp_H\colon\mathfrak h\to H$.
For $X\in\mathfrak g$, we write $\mathrm{Ad}_G(g)X$
for the adjoint action of $g\in G$ on $\mathfrak g$.  A Lie crossed module $(\Phi:H\!\to\!G,\triangleright)$ consists
of a Lie group morphism $\Phi\colon H\to G$ and a smooth left action
$\triangleright$ of $G$ on~$H$ by automorphisms satisfying
\begin{equation}
\Phi(g\!\triangleright\!h)=g\,\Phi(h)\,g^{-1},
\qquad
\Phi(h_1)\!\triangleright\!h_2=h_1h_2h_1^{-1}.
\end{equation}
Its differential
$(\partial:\mathfrak h\!\to\!\mathfrak g,\triangleright)$
is called a differential crossed module.
The kernel $\ker\Phi\subseteq H$ will occasionally be referred to as
the inert subgroup, following the terminology of~\cite{yek}.  For a smooth manifold $M$, we write $\Omega^k(M,\mathfrak g)$ for the
space of smooth $\mathfrak g$--valued $k$--forms on~$M$.
If $\alpha\in\Omega^1(M,\mathfrak g)$ is a connection $1$--form, its
curvature is $F_\alpha=d\alpha+\tfrac12[\alpha,\alpha]$.
Given a differential crossed module
$(\partial:\mathfrak h\!\to\!\mathfrak g,\triangleright)$, a $2$--connection on~$M$ is a pair
$(\alpha,\beta)$ with
$\alpha\in\Omega^1(M,\mathfrak g)$ and
$\beta\in\Omega^2(M,\mathfrak h)$ which satisfies the fake--flatness condition

\begin{equation} F_\alpha+\partial(\beta)=0. \end{equation}

\noindent
Its $3$--curvature is
\begin{equation}
\mathcal H = d\beta + \alpha\triangleright\beta
   \in \Omega^3(M,\mathfrak h).
\end{equation}

\noindent
If $\{A_i\}$ is a finite collection of group elements indexed by an
ordered set, the notation
$\prod_i A_i$ always denotes the product taken in the increasing order
of the indices.
This convention agrees with the orientation of subdivided simplices in
the construction of the multiplicative integral.
For an oriented simplex $\sigma$, the integral
$\int_\sigma$ denotes the usual signed integral of differential forms
according to this orientation.  

The notation used in the article may be summarized as follows:

\begin{center}
\renewcommand{\arraystretch}{1.2}
\begin{tabular}{ll}
$\Delta^n$ & standard $n$--simplex with vertices $v_0,\dots,v_n$\\
$\sigma\colon\Delta^1\to M$ & smooth path in $M$\\
$\tau\colon\Delta^2\to M$ & smooth oriented surface in $M$\\
$(\sigma,\tau)$ & kite (path and surface sharing a base point)\\
$\mathrm{MI}(\alpha|\sigma)$ & $G$--valued multiplicative integral of $\alpha$ on~$\sigma$\\
$\mathrm{MI}(\alpha,\beta|\sigma,\tau)$ & $H$--valued multiplicative integral on a kite\\
$F_\alpha$ & curvature of the $1$--form connection $\alpha$\\
$\mathcal H$ & $3$--curvature of the $2$--connection $(\alpha,\beta)$\\
$\Phi:H\!\to\!G$ & morphism of crossed modules\\
$\triangleright$ & $G$--action on $H$ and $\mathfrak h$\\
\end{tabular}
\end{center}

\subsection{Multiplicative integration on surfaces}

\noindent
We briefly recall the construction of multiplicative
integration as developed by Yekutieli in \cite{yek}.
All analytic estimates, convergence proofs, and geometric background
on piecewise-smooth simplices may be found there and we will include here only essential definitions and results.

Let $M$ be a smooth manifold and $G$ a Lie group with Lie algebra
$\mathfrak g$.
Given a smooth path $\sigma\colon\Delta^1\to M$ and a
$\mathfrak g$-valued $1$-form $\alpha\in\Omega^1(M,\mathfrak g)$, the multiplicative integral
$\mathrm{MI}(\alpha|\sigma)\in G$ is defined to be the limit of iterated
Riemann products.
If $\sigma$ is subdivided into $2^k$ equal subintervals
$\sigma_1,\dots,\sigma_{2^k}$, one sets
\begin{equation}
\mathrm{RP}_k(\alpha|\sigma)
   :=\prod_{i=1}^{2^k}\exp_G\!\Big(\!\int_{\sigma_i}\!\sigma_i^*\alpha\Big),
\end{equation}
where the product is ordered along the orientation of~$\sigma$.  The limit
\begin{equation}
\mathrm{MI}(\alpha|\sigma)
   := \lim_{k\to\infty}\mathrm{RP}_k(\alpha|\sigma)
\end{equation}
exists and depends smoothly on $\sigma$ by Theorem 3.3.19 of \cite{yek}.  It satisfies the composition law
\begin{equation}
\mathrm{MI}(\alpha|\sigma_2\!\ast\!\sigma_1)
   =\mathrm{MI}(\alpha|\sigma_2)\,\mathrm{MI}(\alpha|\sigma_1),
\end{equation}
and reduces in the abelian case to

\begin{equation} \mathrm{MI}(\alpha|\sigma)=\exp_G \bigg(\int_\sigma\alpha \bigg). \end{equation}

\noindent
It follows that $\mathrm{MI}(\alpha|\sigma)$ recovers the usual holonomy of the
connection~$\alpha$ along~$\sigma$.

To define parallel transport over surfaces, we next introduce
a Lie crossed module $(\Phi:H\!\to\!G,\triangleright)$,
consisting of Lie groups $H$ and $G$, a smooth homomorphism
$\Phi\colon H\to G$, and a smooth action
$\triangleright\colon G\to\mathrm{Aut}(H)$
satisfying
\begin{equation}
\Phi(g\!\triangleright\!h)=g\,\Phi(h)\,g^{-1},
\qquad
\Phi(h_1)\!\triangleright\!h_2=h_1h_2h_1^{-1}.
\end{equation}
Let $(\mathfrak h,\mathfrak g,\partial,\triangleright)$ denote the
corresponding differential crossed module.  A $2$-connection on~$M$ with values in
$(\Phi:H\!\to\!G,\triangleright)$
is a pair of differential forms
$(\alpha,\beta)$ with
$\alpha\in\Omega^1(M,\mathfrak g)$ and
$\beta\in\Omega^2(M,\mathfrak h)$ satisfying the fake-flatness condition
\begin{equation}
F_\alpha+\partial(\beta)=0.
\end{equation}
Its $3$-curvature is
$\mathcal H=d\beta+\alpha\triangleright\beta$.

Let $(\sigma,\tau)$ denote a kite in~$M$, i.e.~a pair consisting
of a base path $\sigma\colon\Delta^1\to M$ and a
surface map $\tau\colon\Delta^2\to M$ such that
$\sigma(v_1)=\tau(v_0)$.
For such data, one can define recursively
$H$-valued Riemann products
$\mathrm{RP}_k(\alpha,\beta|\sigma,\tau)$ by subdividing
$\Delta^2$ into $4^k$ smaller triangles and setting
\begin{equation}
\mathrm{RP}_0(\alpha,\beta|\sigma,\tau)
   := \exp_H\!\Big(
        \Psi(\mathrm{MI}(\alpha|\sigma))
        \!\int_{\tau}\!\tau^*\beta
      \Big),
\end{equation}
with higher-order products obtained by ordered multiplication over the
subdivision.
The limit
\begin{equation}
\mathrm{MI}(\alpha,\beta|\sigma,\tau)
   :=\lim_{k\to\infty}
       \mathrm{RP}_k(\alpha,\beta|\sigma,\tau)
\end{equation}
exists and depends smoothly on $(\sigma,\tau)$ by Theorem 8.3.1 of \cite{yek}.  This $H$-valued element may be interpreted as the surface holonomy associated to the $2$-connection $(\alpha,\beta)$.  In the abelian case where $G=\{1\}$ and $H=U(1)$, this reduces to
\begin{equation}
\mathrm{MI}(\alpha,\beta|\sigma,\tau)
   = \exp_{U(1)}\!\Big(\!\int_{\tau}\!\beta\Big),
\end{equation}
recovering the familiar holonomy of a $U(1)$-bundle gerbe.

The following properties hold for multiplicative integration:

\begin{itemize}
\item[(1)] \textbf{Functoriality.}
  The assignments
  $\sigma\mapsto\mathrm{MI}(\alpha|\sigma)$ and
  $(\sigma,\tau)\mapsto\mathrm{MI}(\alpha,\beta|\sigma,\tau)$
  respect concatenation and inversion of paths and surfaces.

\item[(2)] \textbf{Compatibility with the boundary map.}
  The homomorphism of crossed modules satisfies
  \begin{equation}
  \Phi\big(\mathrm{MI}(\alpha,\beta|\sigma,\tau)\big)
    = \mathrm{MI}(\alpha|\partial\tau).
  \end{equation}

\item[(3)] \textbf{Stokes law.}
  For a small $3$-simplex $f\colon\Delta^3\to M$,
  one has (Lemma 8.4.1 of \cite{yek})
  \begin{equation}
  \mathrm{MI}(\alpha,\beta|\partial f)
    = \exp_H\!\Big(\!\int_f\!\mathcal H
        +O(|f|^2)\!\Big).
  \end{equation}
\end{itemize}

\section{Curvature obstruction and the Wess-Zumino phase}

\noindent
The theory of nonabelian multiplicative integration on surfaces provides analytic
definitions of holonomy for connections on crossed modules and
establishes higher-dimensional Stokes theorems.  In particular, Theorem~0.5.4 of \cite{yek} shows that the
$3$-form curvature of a connection-curvature pair
$(\alpha,\beta)$ governs the deviation of the 2D multiplicative integral from being exact.  In this section, we reinterpret this local multiplicative Stokes law
geometrically in terms of higher surface holonomy.  The theorem may be viewed as an infinitesimal curvature-obstruction
relation for higher holonomy, suggesting that differences between
surface holonomies should be governed by the flux of the associated $3$-curvature.  In the abelian case, this leads to a rigorous higher-dimensional phase
relation which reproduces the familiar Wess-Zumino term from string
theory \cite{wess, witten2, green, freed}.  More precisely, we show
that the phase difference between two worldsheet surfaces with common
boundary is determined by the exponential of the integral of the
$3$-form field strength of the $B$-field over an interpolating
three-manifold.  We state the local multiplicative Stokes law in Section 3.1.  We prove the abelian curvature obstruction relation in
Section 3.2 and show in Section 3.3 that it reproduces the standard
Wess--Zumino phase.
We then discuss possible nonabelian generalizations and their relation
to higher gauge theory and generalized global symmetries in Section 3.4.

\subsection{Local multiplicative Stokes law}

\noindent
We begin by recalling the local three-dimensional Stokes theorem for
multiplicative integration
\cite{yek}.  This theorem gives the infinitesimal relation between
surface holonomy and the $3$-curvature of a
$2$-connection taking values in a crossed module.  In the present work, this theorem will serve as the
local geometric input for the later discussion of curvature
obstructions and the Wess-Zumino phase.  Let $(\Phi:H\!\to\!G,\triangleright)$ be a Lie crossed module with
differential crossed module
$(\partial:\mathfrak h\!\to\!\mathfrak g,\triangleright)$.  Furthermore, let $(\alpha,\beta)$ be a smooth $2$-connection on a manifold
$M$.  Recall that the associated $3$-curvature is the
$\mathfrak h$-valued $3$-form
\begin{equation}
\mathcal H
   = d\beta + \alpha\triangleright\beta.
\end{equation}

\noindent
Finally, let
$f\colon\Delta^3\to M$
be a sufficiently small smooth $3$-simplex.
Denote by $\partial_i f$
its oriented $2$--faces and let
\begin{equation}
h_i
   := \mathrm{MI}(\alpha,\beta\mid \partial_i f)
   \in H
\end{equation}
be the corresponding multiplicative surface integrals: the following theorem is the local multiplicative Stokes law.

\medskip

\noindent
\begin{proposition}[Local multiplicative Stokes theorem]
For every sufficiently small smooth $3$-simplex $f\colon\Delta^3\to M$,
the associated surface holonomies satisfy
\begin{equation}\label{eq:local-stokes}
h_1 h_2 h_3 h_4^{-1}
=
\exp_H\!\Big(
   \int_f \mathcal H
   + \epsilon(f)
\Big),
\end{equation}
where
$\epsilon(f)\in\mathfrak h$
satisfies an estimate
\begin{equation}
\|\epsilon(f)\|
   \le C\,|f|^2
\end{equation}
for some constant $C$ independent of $f$ and $|f|$ denotes the diameter of the simplex.
\end{proposition}

\begin{proof}
This is a reformulation of Lemma 8.4.1 of
\cite{yek} expressed in the notation of surface holonomy.  The result follows from the convergence properties of the
multiplicative Riemann products defining
$\mathrm{MI}(\alpha,\beta)$ together with the
Baker-Campbell-Hausdorff estimates established in
Proposition 8.2.1 and Theorem 8.3.1 of \cite{yek}.
\end{proof}

\medskip

\noindent
Equation~\eqref{eq:local-stokes} may be interpreted geometrically as
an infinitesimal curvature obstruction law:
the product of the surface holonomies around the boundary of a small
tetrahedron is governed by the integral of the
$3$-curvature $\mathcal H$ over its interior.  In the abelian case where
$H=U(1)$,
equation~\eqref{eq:local-stokes} reduces to the ordinary exponential
Stokes relation for gerbe holonomy.
This suggests that the local multiplicative Stokes theorem can be developed to give a natural higher analogue of the curvature identities underlying
Wess-Zumino phases in string theory.  We will not attempt to derive the nonabelian phase in this work, but we can show quite straightforwardly that the abelian case produces the usual Wess-Zumino phase.

\subsection{Abelian 3D Stokes theorem and phase relation}

\noindent
We now formulate and prove a global higher-dimensional Stokes
relation under an abelian assumption.
In particular, we assume that the $2$-form $\beta$ and the
$3$-curvature $\mathcal H$ take values in an abelian ideal of
$\mathfrak h$.  In this setting the surface holonomies commute and the local
multiplicative Stokes law globalizes.

\begin{proposition}[Abelian 3D Stokes theorem]
\label{prop:abelian-stokes}

Let $(\alpha,\beta)$ be a smooth $2$-connection on $M$
taking values in a crossed module
\begin{equation}
\partial : H \to G,
\end{equation}
with associated $3$-curvature
\begin{equation}
\mathcal H = d\beta + \alpha \triangleright \beta.
\end{equation}
Assume that $\beta$ and $\mathcal H$ take values in an abelian
ideal of $\mathfrak h$.  Let
\begin{equation}
\Sigma_0,\Sigma_1 : [0,1]^2 \to M
\end{equation}
be oriented surfaces with common boundary and let
\begin{equation}
V : [0,1]^3 \to M
\end{equation}
be a smooth $3$-chain satisfying
\[
\partial V = \Sigma_1 - \Sigma_0.
\]
Then
\begin{equation}
\mathrm{MI}(\alpha,\beta\mid\Sigma_1)
=
\mathrm{MI}(\alpha,\beta\mid\Sigma_0)
\exp_H\!\left(
\int_V \mathcal H
\right).
\end{equation}
In particular, if $\mathcal H=0$, the surface holonomy only depends on the common boundary curve.
\end{proposition}

\begin{proof}

By the local multiplicative Stokes theorem, for every sufficiently small smooth
$3$-simplex $f \subset M$ one has
\begin{equation}
h_1 h_2 h_3 h_4^{-1}
=
\exp_H\!\left(
\int_f \mathcal H + \epsilon(f)
\right),
\end{equation}
where
\begin{equation}
\|\epsilon(f)\|
\le
C\,(\mathrm{diam}(f))^2.
\end{equation}
and the $h_i$ denote the surface holonomies of the four oriented
faces of $f$.  Triangulate the $3$-chain $V$ by sufficiently small simplices
$f_k$.  Applying the local identity to each simplex and multiplying over the
triangulation gives
\begin{equation}
\prod_k
\mathrm{MI}(\alpha,\beta\mid\partial f_k)
=
\exp_H\!\left(
\sum_k \int_{f_k}\mathcal H
+
\epsilon_{\mathcal T}
\right),
\end{equation}
where
\begin{equation}
\epsilon_{\mathcal T}
=
O\!\left(
\max_k (\mathrm{diam}(f_k))^2
\right).
\end{equation}
Because the relevant holonomies lie in an abelian subgroup of $H$,
the internal face contributions cancel pairwise.  Every internal face appears twice with opposite orientation,
contributing factors $h$ and $h^{-1}$.
Only the boundary terms survive, which yields
\begin{equation}
\mathrm{MI}(\alpha,\beta\mid\Sigma_1)
\,
\mathrm{MI}(\alpha,\beta\mid\Sigma_0)^{-1}
=
\exp_H\!\left(
\int_V \mathcal H
+
\epsilon_{\mathcal T}
\right).
\end{equation}
Theorem~8.3.1 of \cite{yek} implies that the multiplicative
Riemann products converge uniformly and are independent of the
chosen subdivision.
Letting the mesh size tend to zero therefore gives
\begin{equation}
\mathrm{MI}(\alpha,\beta\mid\Sigma_1)
=
\mathrm{MI}(\alpha,\beta\mid\Sigma_0)
\exp_H\!\left(
\int_V \mathcal H
\right).
\end{equation}
If $\mathcal H=0$ the exponential term is trivial, so the surface
holonomy only depends on the common boundary.
\end{proof}

\medskip

\noindent
The proof above relies crucially on the abelian assumption: internal face contributions only cancel out because the
relevant holonomies commute.  In the fully nonabelian setting, adjacent face holonomies are related by a twisting from the crossed
module structure and the Baker-Campbell-Hausdorff formula
introduces corrections, complicating the analysis significantly.  The above globalization argument above therefore does not directly extend
to the general case.

\subsection{The Wess-Zumino phase}

\noindent
We now show that the abelian Stokes relation of
Proposition~\ref{prop:abelian-stokes}
reproduces the familiar Wess--Zumino phase appearing in
two-dimensional sigma models.  For the Lie crossed module, choose
\begin{equation}
U(1)\longrightarrow \{1\}.
\end{equation}
In this case, the $2$-connection reduces to an abelian
$2$-form gauge field
\begin{equation}
B\in\Omega^2(M),
\end{equation}
and the associated $3$-curvature is 
\begin{equation}
H=dB.
\end{equation}
The surface holonomy takes values in $U(1)$ and therefore defines
a phase factor.  Applying Proposition~\ref{prop:abelian-stokes} gives
\begin{equation}
\mathrm{Hol}_2(\Sigma_1)
=
\mathrm{Hol}_2(\Sigma_0)
\exp\!\left(
i\int_V H
\right),
\end{equation}
where
\begin{equation}
\partial V=\Sigma_1-\Sigma_0.
\end{equation}
The phase difference between two surfaces with common
boundary is therefore determined by the integral of the $3$-form flux
$H$ over an interpolating $3$-chain.

Let
\begin{equation}
g:\Sigma\to G
\end{equation}
be a sigma model field with a compact Lie group $G$ as target space.  The Maurer-Cartan form
\begin{equation}
\alpha=g^{-1}dg
\end{equation}
satisfies
\begin{equation}
d\alpha+\tfrac12[\alpha,\alpha]=0.
\end{equation}
Using the invariant bilinear form
$\mathrm{Tr}$ on $\mathfrak g$,
one obtains the closed $3$-form
\begin{equation}
H
=
-\frac{k}{24\pi}
\mathrm{Tr}
\bigl(
g^{-1}dg
\wedge
g^{-1}dg
\wedge
g^{-1}dg
\bigr),
\end{equation}
where $k\in\mathbb Z$.  This 3-form generates $H^3(G,\mathbb Z)$.  The corresponding surface holonomy defines a phase
\begin{equation}
\mathrm{Hol}_2(\Sigma)
=
\exp\!\bigl(
iS_{WZ}[g]
\bigr),
\end{equation}
where
\begin{equation}
S_{WZ}[g]
=
-\frac{k}{24\pi}
\int_B
\mathrm{Tr}
\bigl(
g^{-1}dg
\wedge
g^{-1}dg
\wedge
g^{-1}dg
\bigr),
\end{equation}
and $\partial B = \Sigma$.  This is exactly the usual Wess-Zumino action.

Although this definition appears to depend on the choice of
extension $B$, the ambiguity is quantized.
If $B$ and $\widetilde B$ are two extensions of $\Sigma$, then
\begin{equation}
W=B\cup(-\widetilde B)
\end{equation}
is a closed $3$-manifold and
\begin{equation}
S_{WZ}[g]-S_{WZ}[\widetilde g]
=
\int_W H.
\end{equation}
Since
\begin{equation}
\int_W H
=
2\pi k\,n,
\qquad n\in\mathbb Z,
\end{equation}
it follows that
\begin{equation}
S_{WZ}[g]-S_{WZ}[\widetilde g]
\in
2\pi\mathbb Z,
\end{equation}
which is the usual Wess-Zumino quantization condition.  It follows that in the abelian setting, the formalism for multiplicative integration on surfaces reproduces the standard Wess-Zumino phase as the logarithm of a surface holonomy.

\subsection{Nonabelian generalization}

\noindent
The abelian derivation discussed above suggests the possibility
of more general topological actions associated to nonabelian higher
gauge fields.  Let
\begin{equation}
\partial:H \to G
\end{equation}
be a Lie crossed module equipped with a $2$--connection $(\alpha, \beta)$.  Multiplicative integration associates to a suitably
parametrized oriented surface $\Sigma$ a surface holonomy $\mathrm{Hol}_2(\Sigma)\in H$.  In the nonabelian setting, the surface holonomy itself no longer
defines a scalar phase directly.  Nevertheless, gauge-invariant
quantities may be obtained by applying suitable representations,
characters, or class functions on $H$ in close analogy with Wilson
loop observables in ordinary gauge theory.

The local multiplicative Stokes theorem of Yekutieli suggests that the
variation of these higher surface observables should be governed by
the associated $3$-curvature of the $2$-connection but in the
fully nonabelian setting one expects additional ordering, transport,
and twisting effects arising from the crossed-module structure and the
underlying higher parallel transport.  The abelian phase relation proved above should therefore be viewed
as the simplest example of a more general higher holonomy
phenomenon which is beyond the scope of this work to derive.  Developing a fully rigorous global nonabelian phase
formula remains an interesting open problem.  From the physical point of view, these constructions are related to higher Wilson surface observables and to theories with
generalized global symmetries, where extended objects couple
naturally to higher gauge fields.  The multiplicative integration
formalism therefore provides a promising analytic framework for the
study of nonabelian higher gauge theory and its associated topological
terms.

\section{Conclusion}

\noindent
In this article we have examined the geometric interpretation of
Yekutieli's multiplicative integration on surfaces in the context of
higher gauge theory.  Starting from the local multiplicative Stokes
theorem for $2$-connections taking values in crossed modules, we interpreted the
associated $3$-curvature as the local obstruction governing the
variation of the surface holonomy.  Under an abelian assumption, we can then establish a globalization of this result which expresses the difference between
the surface holonomies of two surfaces with common boundary in terms
of the exponential of the integrated $3$-curvature over an
interpolating $3$-chain.  This provides a rigorous higher-dimensional
analogue of the familiar relation between curvature and holonomy in
ordinary gauge theory.  We then showed that this formula reproduces
the standard Wess-Zumino phase relation.  In particular, the
Wess--Zumino action arises naturally as the logarithm of the surface
holonomy associated to an abelian higher gauge field and the usual
quantization condition follows from the integrality of the underlying
cohomology class.  From this perspective, multiplicative integration on surfaces
provides a concrete differential geometry framework for understanding
the origin of the Wess-Zumino phase.

There are several directions for future
investigation.  The local nonabelian multiplicative Stokes theorem
strongly suggests the existence of a corresponding nonabelian version of the phase relation which we derived in the abelian case.  However, a complete treatment of the ordering and twisting associated to crossed
modules is beyond the scope of this work.  Establishing such
a theory would provide a natural route toward genuinely nonabelian
higher Wess-Zumino phases with potential applications in string theory, condensed matter and generalized symmetries.  More generally, the surface holonomies arising from multiplicative
integration provide natural candidates for higher Wilson surface
observables in higher gauge theory.  It would be interesting to
investigate their role in theories with generalized global symmetries,
as well as possible connections with anomaly inflow, Chern--Simons
transgression, and related higher-gauge-theoretic constructions.  The abelian phase relation derived above bears a formal resemblance to transgression formulas appearing in Chern–Simons theory and anomaly inflow. In both situations a boundary phase is controlled by the integral of a higher-degree curvature form over a bulk region: nonabelian multiplicative integration on surfaces could provide a natural higher-gauge-theoretic framework for such constructions.  Finally, the comparison with the transport $2$-functor approach of
Schreiber and Waldorf suggests that multiplicative integration could also provide a concrete
analytic realization of higher parallel transport.  Further study of
the relationship between these viewpoints may help clarify the
differentia geometry foundations of higher gauge theory and its
applications in mathematical physics.

\section*{Acknowledgments}

\noindent
The author thanks Severin Bunk and Amnon Yekutieli for useful discussions.  This research was partly conducted whilst the author was visiting the Okinawa Institute of Science and 
Technology (OIST) through the Theoretical Sciences Visiting Program (TSVP).

\section*{Data availability statement}

\noindent
No datasets were generated or analysed during the current study. All calculations and derivations are presented explicitly in the text.

\section*{Conflict of interest}

\noindent
The author declares that there are no conflicts of interest.

\end{document}